\documentstyle [12pt]{article}

\begin{document}
\title{The Friedmann integrals and physical vacuum \\
in the framework of macroscopic extra dimensions}
\author{A.D. Chernin\\
Sternberg Astronomical Institute, Moscow University,\\
Tuorla Observatory, University of Turku\\
 }
\date{}
 \maketitle
\vspace{0.5cm}

{\bf The structure and origin of the Friedmann integrals are analyzed within
the  framework of large extra dimensions proposed by Arkani-Hamed et all.
(1998). It is demonstrated that the integrals might emerge from
extra-dimension physics and reveal its signature.
In the case of two extra dimensions, the
integrals are expressed via the product $M_* R$ of the only fundamental energy
$M_*$ and the size $R $ of the extra dimensions. The cosmic
vacuum density turns out to be $\sim R^4$, in this case. The effective
cut-off in the spectrum of the quantum zero-point oscillations at the
frequency $\omega_V \sim 1/R$ is assumed to be associated with the epoch
when the Hubble radius becomes larger than $R$. This occurs at temperatures
$ \sim M_*$ in the first three picoseconds of the cosmic expansion.
}
\newpage

\section{Introduction}

The Friedmann cosmology is a successful theory that
predicted the cosmic expansion 80 years ago. It is consistent
with all the bulk of current observational results, including
acceleration of the expansion and the existence of cosmic vacuum
recognized recently
with the SN type Ia observations [1].  Any generalization or extension of
standard cosmology must include the Friedmann theory as a well-working
particular or asymptotic case. On the other hand, if the Friedmann theory
is really a limit of a more general theory, it may keep some features of
such a theory. These features can be mirrored in the
free (empirical) parameters of the theory which are the Friedmann integrals.

The Friedmann integrals are associated with the energy
content of the Universe and represent four major forms of cosmic
energy: cosmic vacuum, dark matter, baryons and radiation.
The integrals appear as a result of integration
of the Friedmann `second' equation which is essentially the
internal energy conservation law for each of the energy forms,
and they determine the structure of the Friedmann `first' equation
which is the mechanical energy conservation law, if to use the notions
of the Newtonian physics.

With the use of the current dataset, the numerical values of the
Friedmann integrals can be estimated, and they have proven to
be equal to each other, on the order of
magnitude. If this is not a purely arithmetical coincidence, the near
equality of the Friedmann integrals is a time-independent symmetry
relation that puts cosmic vacuum in correspondence with the non-vacuum cosmic
energies [2]. Symmetries that do not concern space-time geometry are usually
called internal symmetries, in particle physics; we face now cosmic internal
symmetry (CISY) that exists as long as the observed forms of cosmic energy
are present in nature.

In this note, I discuss the framework of macroscopic extra
dimensions proposed in [3] as a multidimensional extension of the Friedmann
cosmology; I examine how the Friedmann integrals and CISY may reflect and
reveal the basic physics of extra-dimensions, if they really exist.
The origin of cosmic vacuum,
which is actually the same object as vacuum of particle physics,
is one of the major aspects of the problem under consideration.

\section{Four integrals}

The Friedmann integrals are given by a common relation
\begin{equation}
A = [ (\rho a^{3(1+w)}]^{\frac{1}{1 + 3w}},
\end{equation}
\noindent
where $a(t)$ is the 3D curvature radius and/or scale factor;
$\kappa = 8 \pi G/3$ is the Einstein gravitational constant;
$w = p/\rho$ is the pressure-density ratio for each of the energy forms:
$w = -1$ for vacuum (V),  $w = 0$ for dark (D) matter and baryons (B),
$w = 1/3$ for radiation (R). Hereafter $c = 1$.

The integrals enter the basic equation for the cosmological expansion
in a rather symmetrical way:
\begin{equation}
{\dot a}^2 = (A_V/a)^{-2} + A_D/a + A_B/a + (A_R/a)^2  - k.
\end{equation}
\noindent
Here $k = 1, 0, -1$ for close, flat and open models, respectively.

The numerical values of the integrals can be estimated with the use of the
observational dataset
on the cosmic densities, the age of the Universe and the Hubble
constant. The result is as follows [2]:
\begin{equation}
\begin{array}{l}
A_V = (\kappa \rho_V)^{-1/2} \sim 10^{61} M_{Pl}^{-1}, \\

A_D = \kappa \rho_D a^3 \sim10^{61} M_{Pl}^{-1}, \\

A_B = \kappa \rho_B a^3 \sim 10^{59} M_{Pl}^{-1}, \\

A_R = (\kappa \rho _R)^{1/2} a^2  \sim 10^{59} M_{Pl}^{-1}. \\
\end{array}
\end{equation}
\noindent
Here
 $\rho_V, \rho_D, \rho_B, \rho_R$ are the densities of vacuum,
 dark matter, baryons, and radiation, correspondingly.
The units are used in which $c = \hbar = 1$; $G = M_{Pl}^{-2}$, and the
Planck mass $M_{Pl} = 1.2 \times 10^{19}$ GeV.

As we see, the four integrals are equal to each other within two orders
of magnitude, and their equality can be treated as a symmetry relation. This
symmetry is not exact; CISY is violated at the level of a few percent,
on the logarithmic scale. Its physical nature seems to be not too
mysterious: it may be understood as a result of the cosmic
freeze-out process in the early Universe [2].

\section{Fundamental energy scales}

The standard freeze-out model for the particle-antiparticle annihilation
 assumes that the dark matter particles are
thermal relics of the hot initial stages of the cosmological expansion.
With the use of a fairly transparent version of this
model [4],  one can show that the near equality of the Friedmann
integrals for vacuum, dark matter and radiation is a direct outcome
of the freeze-out process at the first three picoseconds [2]. One can
also find with the freeze-out model that the Friedmann integrals are
expressed in the terms of
two fundamental energy scales which are the Planck mass and the
electroweak mass $M_{EW} \sim 1$ TeV:
\begin{equation}
A_V \sim A_D \sim A_R \sim  (\bar M_{Pl}/M_{EW})^4 \bar M_{Pl}^{-1}
\sim g^3 (M_{Pl}/M_{EW})^4  M_{Pl}^{-1}
\sim 10^{61 \pm 1} M_{Pl}^{-1}.
\end{equation}
\noindent
Here $\bar M_{Pl} = g M_{Pl}$ is the reduced Planck mass, $g \simeq 0.1 - 0.3$.

With the same accuracy and in accordance with Eqs.(1,3) one has for
the constant vacuum density (cf. also [4]):
\begin{equation}
\rho_V \sim g^8 (M_{Pl}/M_{EW})^8 M_{Pl}^4 \sim 10^{-122 \pm 2} M_{Pl}^4.
\end{equation}
\noindent

The Friedmann integral for baryons is not included in the freeze-out model;
perhaps it may be obtained with the current models of electroweak
baryogenesis (this topic is reviewed in [5]). Accidentally or not,
the Baryonic Number can also be expressed numerically in the terms of the two
fundamental energy scales:
\begin{equation}
B \sim (\bar M_{Pl}/M_{EW})^{2/3} \sim 10^{10}.
\end{equation}

The big dimensionless ratio $ \bar M_{Pl}/M_{EW} \sim 10^{15}$ proves to
be the key quantity that determines the Friedmann integrals and the
density of cosmic vacuum, according Eqs.(4,5). In particle
physics, the nature of the huge gap between $M_{Pl}$ and $M_{EW}$ is
not explained and known as the hierarchy problem. We see now that
cosmology is highly sensitive to this problem as well.

\section{Extra dimensions}

The framework of large extra dimensions [3] provides a new prospective  for
the hierarchy problem. It may also suggest a new understanding of the
structure of the Friedmann integrals.

It is assumed in [3] that there is one and only one fundamental energy
scale in nature, and this scale $M_*$ is close to the electroweak scale
$M_{EW}$. As for the Planck mass, it is due to the extra
dimensions of space:
\begin{equation}
M_{Pl} \sim (M_{*} R)^{n/2} M_{*}.
\end{equation}
\noindent
Here $n, R$ are the number of the spatial extra dimensions and their size,
which is proposed to be the same for all of them. It is also argued in [3]
that the case $n = 2$ is the most promising;
in this case the size of two extra dimensions
\begin{equation}
R \sim 0.1 \;\;cm, \;\;\; n = 2.
\end{equation}

In accordance with Eq.(7), the big dimensionless  ratio
$ M_{Pl}/M_{EW} $ is replaced now with the product $ M_* R$. Treating
the extra dimension framework as a multi-dimensional extension of the
Friedmann theory, one may obtain the Friedmann integrals in the terms
of the new fundamental energy and the size of extra dimensions:
\begin{equation}
A \sim (M_{EW} R)^{(3/2)n} M_{EW}^{-1}.
\end{equation}
\noindent
In the case of two extra dimensions, on has from this:
\begin{equation}
A \sim (M_{EW} R)^{2} R, \;\;\;\; n = 2.
\end{equation}

Then one finds for vacuum density a general relation
\begin{equation}
\rho_V  \sim (M_{EW} R )^{-2n} M_{EW}^{4},
\end{equation}
and a particular relation in the case of two extra dimensions:
\begin{equation}
\rho_V \sim R^{-4}; \;\; n = 2.
\end{equation}

Finally, one finds for the Baryonic Number:
\begin{equation}
B  \sim (M_{EW} R )^{n/3}.
\end{equation}
\noindent
Only two extra dimensions provide the correct value of $B$:
\begin{equation}
B \sim (M_{EW} R)^{2/3} \sim 10^{10}; \;\;\;\; n = 2.
\end{equation}

\section{Discussion}

As we see, the multi-dimensional extension of the standard cosmology sheds
new light on the origin of the free parameters of the Friedmann theory.
These parameters -- the Friedmann integrals -- might emerge
from physics of extra dimensions and carry their signature.
According to Eqs.(9-14), this physics
is explicitly imprinted in the structure of the Friedmann integrals:
the integrals turn out to be expressed in the terms of two basic quantities
of extra-dimension physics -- the fundamental energy scale and the size
of extra dimensions. The two quantities appear in a dimensionless product
$M_{EW} R$, which is of the order of $ 10^{16}$ for the case of two extra
dimensions. It is not surprising that this new dimensionless number is similar
to the mass ratio $M_{Pl}/M_{EW}$ involved in the cosmological freeze out
(see Sec.2). Actually, we have now a reformulated hierarchy problem:
instead of one of the two energy scales, a new spatial scale
appears in the framework [3].

What is surprising is that the new expression (Eq.12) for the vacuum density
is free from any signs of hierarchy problem: the density is
 a power of the extra-dimension size only, $\rho_V \sim R^{-4}$,
in case of two extra dimensions.

One may assume, as it has been done earlier
not once, that the finite value of the vacuum density is a result of
an effective  cut-off in the frequency spectrum of the zero-point
oscillations of quantum fields. If so, the cut-off frequency is
$\omega_V \sim 1/R$, in the case of two extra dimensions. It means that
the vacuum energy on the brane is due to the low-frequency band of the
oscillations only, $\omega \le \omega_V \sim 1/R$. The wave lengths of
zero-point oscillations are all larger than the size of the extra dimensions:
$\lambda_V \ge 1/\omega_V \sim R \sim 1$ mm.

It seems instructive that the energy density in the Casimir effect is the
same power function of a size: $\rho_C \sim d^{-4}$. Here $d$ is the
distance between two parallel plates, or concentric spheres, etc.
which is much
smaller than the other sizes $L$ in the experiment. Recent measurements of the
Casimir force between a cantilever and a plate see in [6]
(the separation was between 0.5 and 3 microns; the accuracy  15\%).
This analogy, especially in the case of two plates, suggests that the
frequency cut-off and the origin of the finite energy density can really
be due to a small ratio of the sizes involved, like $d/L << 1$.

In the context of the early Universe, a small size ratio appears when the
the Hubble radius exceeds the size $R \sim 1$ mm of extra dimensions.
It occurs in the era when the cosmic temperature is about $M_* \sim 1$ TeV,
and the cosmic age is a few picoseconds. According to the consideration
above, this event might lead to the effective cut-off in the
spectrum of zero-point quantum oscillations. One can assume, that since
that epoch, the vacuum density on the brane is finite and constant, and it
keeps then its value the same forever. The
 Friedmann cosmology with its basic parameters and internal symmetry
becomes valid since that era.

Three other almost simultaneous events might happen at the same era of a
few picoseconds: electroweak symmetry breaking, electroweak baryogenesis,
and freeze out of the dark matter annihilation. It may hardly be a simple
coincidence, and one common physical mechanism is rather behind
all the four cosmic events.
It seems remarkable that the quantitative characteristics of the four
events are directly associated  with the fundamental energy scale $ M_*$
and the size of extra dimensions $ R$, according to Eqs.(9-14).

The earlier pre-Friedmann evolution of the Universe is essentially
multi-dimensional, and the framework [3] provides
promising grounds for the study of this evolution.
In its turn, this earlier evolution might
start at the epoch when the cosmic age was $t_M \sim 1/M_* \sim 10^{-27}$
sec. This is 16 orders of magnitude larger than the Planck time.

Might it be that the history of the Universe and the very time take
start at $t_M$? Indeed, if energies
considerably larger than $M_*$ are not possible in nature at all,
no temporal structures finer than $\sim 1/M_*$ can physically be resolved.
An answer to the question can be expected from the coming experiments
at the new big colliders. If it occurs that any considerable excess of
the particle energy above $\sim M_*$ is prevented by the generation of new
particles, black holes, gravitational waves, etc., it will mean that the
principal upper energy limit and the
principal lower time limit do exist in nature.

On the other hand, further submillimeter laboratory experiments with
the Casimir effect, deviations from the gravity inverse-square law, etc.
may identify the size of spatial extra dimensions or at least their upper
limit. As is discussed in [7], not 1 TeV, but rather 10 - 30 TeV
is an appropriate value for $M_*$ in the case of two dimensions; if so,
a constant dimensionless factor $q = M_* /M_{EW}$ must be introduced to
the relations above, where $1 \le q \le 30$. Then one will have (with the
account also of the reduced Planck mass):
$A \sim (g  q^{2}  M_{EW} R)^{3} M_{EW}^{-1}$ and
$\rho_V \sim g^{-4} q^8 R^{-4}$, where $R$ is about 1 mm as above.
The larger value of the fundamental mass corresponds to the size of
extra-dimensions about 1 - 10 microns, which is, by the way, near the
spatial separation in the experiment [6].

Note finally that the framework of large extra dimensions [3] offers a simple
explanation to the phenomenon of non-exact symmetries (like, for
instance, chiral symmetry) and small symmetry
breaking. This explanation addresses the existence of other branes in the
bulk world: if symmetry breaking originates on a distant brane, it should
be small on our brane, because what happens far away affects our brane only
weakly. In this way, one may speculate that CISY is non-exact due to the weak
influence of a distant brane or a separate fold of our own brane.

\vspace{1.5cm}

{\bf References}
\vspace{0.5cm}

[1] Riess A G et al. Astron. J  116 1009 (1998);
Perlmutter S et al. Astrophys. J  517 565 (1999)

[2] Chernin A D Physics-Uspechi  44 1099 (2001);
    New Astron. 7 113 (2002)

[3] Arkani-Hamed N, Dimopoulos S, Dvali G Phys. Lett. B429 263 (1998);
    Antoniadis I, Arkani-Hamed N, Dimopoulos S, Dvali G Phys. Lett.
    B436 257 (1998)

[4] Arkani-Hamed N,  Hall L J, Kolda C,  Murayama H Phys. Rev. Lett. 85
4434 (2000)

[5] Rubakov V A, Shaposhnikov M E Physics-Uspechi 166 493  (1996)

[6] Bressi G et al. Phys. Rev. Lett. 80 041804 (2002)

[7] Rubakov V A Physics-Uspechi 171 913 (2001)

\end{document}